\documentclass[aps,
		prl,
		reprint,
		twocolumn,
		superscriptaddress,
		shortbibliography,
		footinbib,
		floatfix,
		notitlepage
		]{revtex4-1}
\pdfoutput=1
\usepackage{amsmath,amssymb,amsfonts}

\usepackage{natbib}
\usepackage[T1]{fontenc}
\usepackage[latin1]{inputenc}

\usepackage[dvipsnames]{xcolor}

\usepackage{hyperref}
\hypersetup{colorlinks=true,citecolor=Red,urlcolor=Blue}

\usepackage{mathrsfs}
\usepackage{bbm}
\usepackage{slashed}
\DeclareSymbolFont{rsfs}{U}{rsfs}{m}{n}
\DeclareSymbolFontAlphabet{\mathrsfs}{rsfs}
\usepackage[mathscr]{eucal}	
\usepackage[normalem]{ulem}
\usepackage{verbatim}

\usepackage{graphicx}

\newcommand{\be}{\begin{equation}}
\newcommand{\ee}{\end{equation}}
\newcommand{\bi}{\begin{itemize}}
\newcommand{\ei}{\end{itemize}}
\newcommand{\bea}{\begin{eqnarray}}
\newcommand{\eea}{\end{eqnarray}}
\newcommand{\bra}[1]{\langle\,#1\,|}          
\newcommand{\ket}[1]{|\,#1\,\rangle}          
\newcommand{\brat}[1]{\langle\,#1|}          
\newcommand{\kett}[1]{|#1\,\rangle}          
\newcommand{\ud}{\mathrm{d}}		
\newcommand{\Eins}{\mathbbmss{1}}

\newcommand{\LCm}{{\scriptscriptstyle -}} 
\newcommand{\LCp}{{\scriptscriptstyle +}}

\newcommand{\rt}{\mathsf{t}}

\begin{document}
\title{Renormalisation group flow of the Jaynes-Cummings model}
\author{Anton Ilderton}
\email{anton.ilderton@plymouth.ac.uk}
\affiliation{Centre for Mathematical Sciences, University of Plymouth, Plymouth, PL4 8AA, UK}
\begin{abstract}	
The Jaynes-Cummings model is a cornerstone of light-matter interactions.
While finite, the model provides an illustrative example of renormalisation in perturbation theory. We show, however, that exact renormalisation reveals a rich non-perturbative structure, and that the model provides a physical example of a theory with a chaotic coupling trajectory and multi-valued beta-function.
We also construct an exact Wilsonian-like renormalisation group flow for the effective scattering matrix, and show how multi-valued features arise in the flow.  Our results shed light on non-perturbative aspects of renormalisation and on the structure of the Jaynes-Cummings model itself.
\end{abstract}
%

\maketitle
The Jaynes-Cummings model, describing a single electromagnetic mode interacting with a two-level atomic system~\cite{Jaynes:1963zz}, underlies light-matter interactions~\cite{JCEXT2004,JCEXPT2010}, cavity QED~\cite{Walther2006} and circuit QED~\cite{Deppe2008}. 

{Fundamental aspects of the model continue to attract attention~\cite{Nazir1,DiStefano2019}.} We consider here the renormalisation of the Jaynes-Cummings model (JCM). Renormalisation is often introduced, in quantum field theory, as a necessary tool {for removing ultra-violet (UV) divergences which arise through virtual particle loops in perturbation theory}. At each order of perturbation theory, the parameters of the theory are adjusted to match some observational input, which removes divergences and, by fixing the parameters, makes the theory predictive. From this perspective it becomes clear, as is known but not as frequently discussed, that even UV-finite theories require a `finite renormalisation' of their free parameters~\cite{Weinberg:1995mt,Delamotte:2002vw}. 

The JCM is no exception. It requires, as we will see, a finite renormalisation at each order of perturbation theory, analogous to coupling renormalisation in the loop expansion of QED. {This goes through as one might expect, but is by definition limited to the small-coupling regime. As motivation to go beyond this, we note that the $c$ and $a$-theorems~\cite{Zamolodchikov:1986gt,Cardy:1988cwa,Anselmi:1997am,Komargodski:2011vj,Nakayama:2013is} demonstrate the existence of monotonic functions of the renormalisation group (RG) flow, from which it is inferred that periodic or chaotic coupling trajectories are forbidden -- however, exactly solvable models show that renormalisation can lead to exotic behaviour including chaos and limit cycles~\cite{Wilson,Glazek:2002hq,LeClair2004,LeclairRussianDoll,Hammer:2011kg,Curtright-RG}.}
 
We will show here that carrying out renormalisation of the apparently simple JCM \textit{non}-perturbatively reveals a surprising depth of structure. We will see that one can construct a beta-function, describing the RG flow of the coupling, which is multivalued~\cite{Curtright-RG}, with the direction of the flow reversing when encountering branch points. Despite this, monotonic functions of the flow exist.  {As such we show that the} JCM provides a physical, and exact, example of the fact that exotic coupling trajectories are not ruled out by the existence of monotonic flow functions.  

 {We will find}  {an unusual physical consequence of the multi-valued flow, namely that} more than one renormalisation condition is needed to fix the single coupling in the JCM and make the theory predictive: we will show how to resolve this. We will also consider the analogue of a Wilsonian effective action approach~\cite{Gies:2006wv,Delamotte:2007pf} to renormalisation of the JCM, constructing an exact RG flow for the effective scattering matrix, and show how a single-valued flow  {can be} compatible with a multi-valued coupling.

\paragraph*{The Jaynes-Cummings Hamiltonian.}
%
The JCM couples a single electromagnetic mode, frequency $\omega$, to a two-level atomic system, ground state $\kett{\downarrow}$ and excited state $\kett{\uparrow}$, with energy gap $\omega_a$. The Hamiltonian is $H=H_0 + g V$ for coupling $g$, where
\be
	H_0 = \omega a^\dagger a + \omega_a \tau_3 \;, \qquad V = a^\dagger \tau_\LCm + a \tau_\LCp \;,
\ee
in which the electromagnetic mode ladder operators $a$, $a^\dagger$ obey $[a,a^\dagger]=1$ as usual, and the $\tau$ operators may be represented as 
$\tau_\LCp = \kett{\uparrow}\brat{\downarrow}$,\,  $\tau_\LCm = \kett{\downarrow}\brat{\uparrow}$ and
$\tau_3 = [\tau_\LCp,\tau_\LCm]/2$.  {The time-evolution operator $U(t)$ in the interaction picture is known exactly~\cite{Stenholm,Shore1993,Bina2012}: taking for simplicity the `resonance limit' in which the detuning $\omega_a - \omega$ (the difference between the photon and atomic energies) is zero, the time evolution operator is, for ${\sf n}:= a^\dagger a$~\cite{FN2} (see also Appendix A),
\be\label{SMdef}
\begin{split}
	U(t) =
	\cos gt \sqrt{\sf{n}+1}  \kett{\uparrow}\brat{\uparrow} 
	+\cos gt \sqrt{\sf{n}} \kett{\downarrow}\brat{\downarrow}  \\
	-ia \tau_\LCp \frac{\sin gt \sqrt{ \sf{n}}}{\sqrt{ \sf{n}}}   
	-i\frac{\sin gt \sqrt{ \sf{n}}}{\sqrt{ \sf{n}}} a^\dagger  \tau_\LCm \;.
\end{split}
\ee
We will re-introduce the detuning later.}
%
\noindent\paragraph*{Renormalisation.}
\noindent The basic observables in the JCM are transition amplitudes between states of form $\ket{j,\updownarrow}$ containing some number of photons $j$, and with the atom in one of its two states. Defining
\be\label{amp1}
	A_j(t):= i\bra{j+1,\downarrow} U(t) \ket{j,\uparrow} \;,
\ee
the simplest observable is the probability $|A_0(T)|^2$ for the atom to decay from its excited state, emitting a photon, after some time $T$. A measurement of this probability, $\mathbb{P}_\text{obs}$, can be used as a renormalisation condition to determine (from the form of (\ref{SMdef})) $g_0 := g T$, and so $g$. To make a more specific analogy with field theory, suppose that the coupling is switched off after some time $t=T$. Then $U(T)$ is the $S$--matrix, which depends only on the dimensionless coupling $g_0$, the analogue of the charge in the QED $S$-matrix.

We start by making contact with perturbation theory. This amounts to evaluating (\ref{amp1}) in powers of the `bare' coupling $g_0$. This expansion has a Feynman-diagram analogy in QED, since the JCM  {interaction vertex describes the emission/absorption of a single photon from an atomic (matter) state, mirroring the three-point vertex of QED}. The `tree level' contribution to the decay probability $|A_0(T)|^2$ is $g_0^2$. We would therefore identify $g_0 = \sqrt{\mathbb{P}_\text{obs}} \equiv g_r$, the physical, or renormalised, coupling. At the next order of perturbation theory, corresponding to 1-loop in QED, one finds that $A_0(T) = g_0 - g_0^3/6$, and so $g_0$ must be adjusted to ensure that the calculated and measured observables still agree.  {Following the usual procedure, see e.g.~\cite[\S2]{Delamotte:2002vw}, we write $g_0$ as a power series in $g_r$, so $g_0 = g_r + \lambda_1 g_r^3 + \ldots$, and repeat the perturbative calculation. The renormalisation condition at order $g_r^3$ then uniquely determines $\lambda_1=1/6$, and so} $g_0$ becomes
\be\label{serien}
	g_0 = g_r + \frac{1}{6}g_r^3 + \mathcal{O}(g_r^5) \;. 
\ee
At each subsequent order of perturbation theory, the renormalisation condition  {similarly uniquely} determines the relationship between the bare and renormalised couplings. The theory is then renormalised to that order in perturbation theory, all as expected. We turn now to exact results and non-perturbative renormalisation.

\begin{figure}[t!]
	\includegraphics[width=\columnwidth]{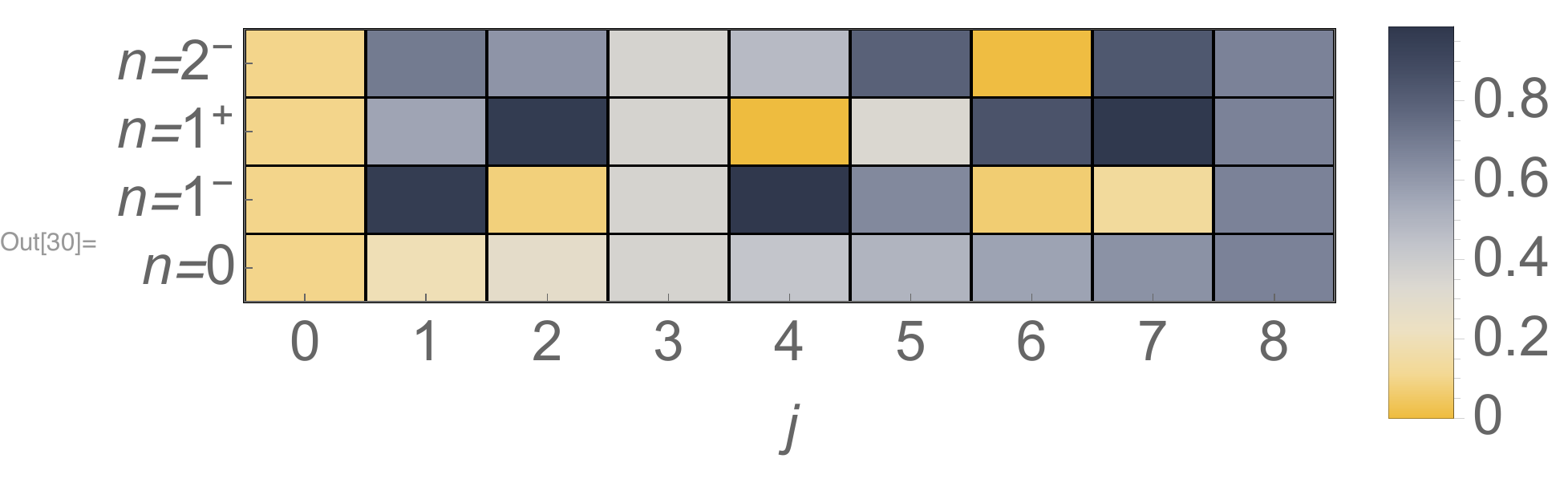}
	\caption{\label{FIG:SPEK} Scattering probabilities $\mathbb{P}_j$ (columns) given by different branches of the solution to the renormalisation condition at $\mathbb{P}_\text{obs}=1/10$. All branches (rows, with $n$ and $\pm$ labelling the choices in (\ref{gren})) yield the same $\mathbb{P}_0 = \mathbb{P}_\text{obs}=1/10$ by construction, but give different $\mathbb{P}_j$ for $j>0$. The fact that the rows differ shows that different branches define different theories.}
\end{figure}
  {From (\ref{SMdef}) the} exact amplitudes $A_j(t)$ are simply
	\be\label{def:an}
	A_j(t) = \sin g t\sqrt{j+1} \;,
\ee
and the renormalisation condition is $g_r = \sin g_0$. This condition has an \textit{infinite} number of solutions corresponding to the possible branches of arcsin,
\be\label{gren}
	g_0  =  \text{arcsin}_n \, \big(\pm \sqrt{\mathbb{P}_\text{obs}}\big) \;,
\ee
in which $\arcsin_n(x) = n\pi + (-1)^n \arcsin (x)$ for $n\in\mathbb{Z}$. 
The choice of branch has physical consequences, as can be seen from the `spectrum' of scattering probabilities $\mathbb{P}_j := |A_j(T)|^2$.  {As shown in Fig.~\ref{FIG:SPEK}, each branch defines a different theory with a different spectrum.}  {Hence we have found that} the renormalisation condition is not enough, \textit{non}-perturbatively, to identify the coupling and make the theory predictive. This situation is unusual, and we will return to it below.

%
%
\paragraph*{Multivalued RG flows.}
We can recast the above discussion in terms of a $\beta$-function. Define the `renormalisation time' $\rt := \log t/T$, then the $\beta$-function $\beta(g_r):= \ud g_r(\rt)/\ud \rt$ describes the evolution of the coupling with respect to time~\cite{Frasca1,Frasca2}, such that $g_r({\sf 0})$ is to match the renormalised coupling above. Again beginning perturbatively,  {we can find the 1-loop $\beta$-function by} inverting the series (\ref{serien}) for $g_r$, differentiating and transforming back: 
\be
	\beta_{1\text{-loop}}(g_r) := g_r - \frac{1}{3}g_r^3 \;.
\ee
From a small $g_r>0$, the coupling (seemingly) flows toward an IR fixed point at $g_r=\sqrt{3}$. This is however outside the perturbative regime, so  {we resum the perturbative series and construct the} all-orders $\beta$-function
\be\label{alla1}
	\beta_0(g_r) := \sqrt{1-g_r^2}\arcsin(g_r) \;,
\ee
 {which corresponds to the $n=0$ branch of (\ref{gren}).} For a flow beginning at $g_r\gtrsim 0$, $\beta_0$ is positive and so $g_r$  {flows toward the turning point at $g_r=1$. At this point, the square root and arcsin in $\beta_0$ switch branch, as they must to account for (\ref{gren}); the $\beta$-function then switches sign and $g$ decreases back toward $-1$, where $\beta$ switches sign again, and so on. As such, because the $\beta$-function is multivalued, the flow continues through the turning points (see~\cite{Curtright-RG} for other examples).}  {We find that, after} encountering $n$ turning points, the exact $\beta$-function is given by  
\be\label{min-beta}
	\beta_n(g_r) = (-1)^n \sqrt{1-g_r^2}\arcsin_n(g_r) \;.
\ee
\begin{figure}[t!]
	\includegraphics[width=0.78\columnwidth]{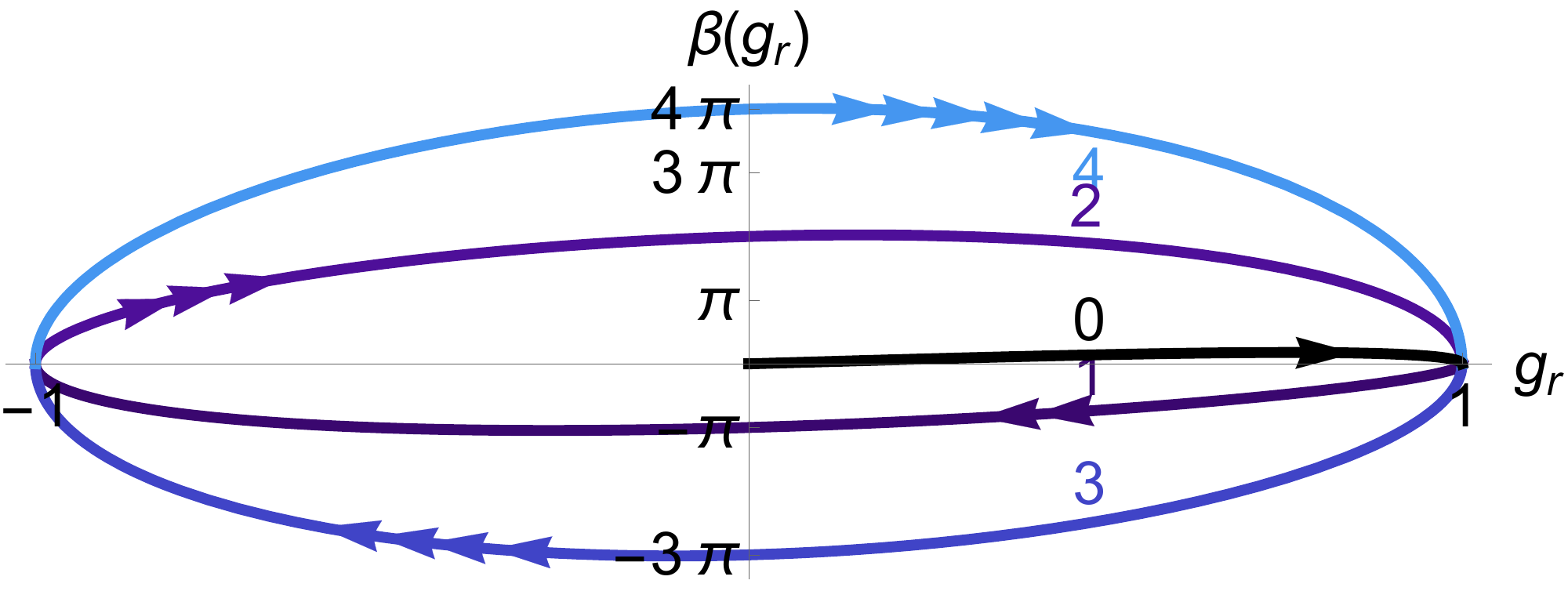}
	\caption{\label{FIG:BETA} Starting from $g_r \gtrsim 0$, the coupling flows toward $g_r=1$, where the $\beta$-function (\ref{min-beta}) switches branch, and the flow turns toward $g_r=-1$, and so on. The flow is shown after encountering $n$ turning points, for $n$ from 0 to 4.}
\end{figure} 
\noindent The flow is illustrated in Fig.~\ref{FIG:BETA}, which shows that the turning points bound the coupling to obey $|g_r|\leq 1$ (as it must from (\ref{def:an})). This means that the fixed point at $g_r=\sqrt{3}$, inferred from 1-loop perturbation theory, can never be reached.  {Thus our results provide} a greatly simplified analogy of the Landau pole in QED, inferred from perturbation theory, but which lies in an inaccessible part of parameter space~\cite{Gockeler:1997dn,Gies:2004hy}.

\begin{figure}[t!!]
	\includegraphics[width=0.93\columnwidth]{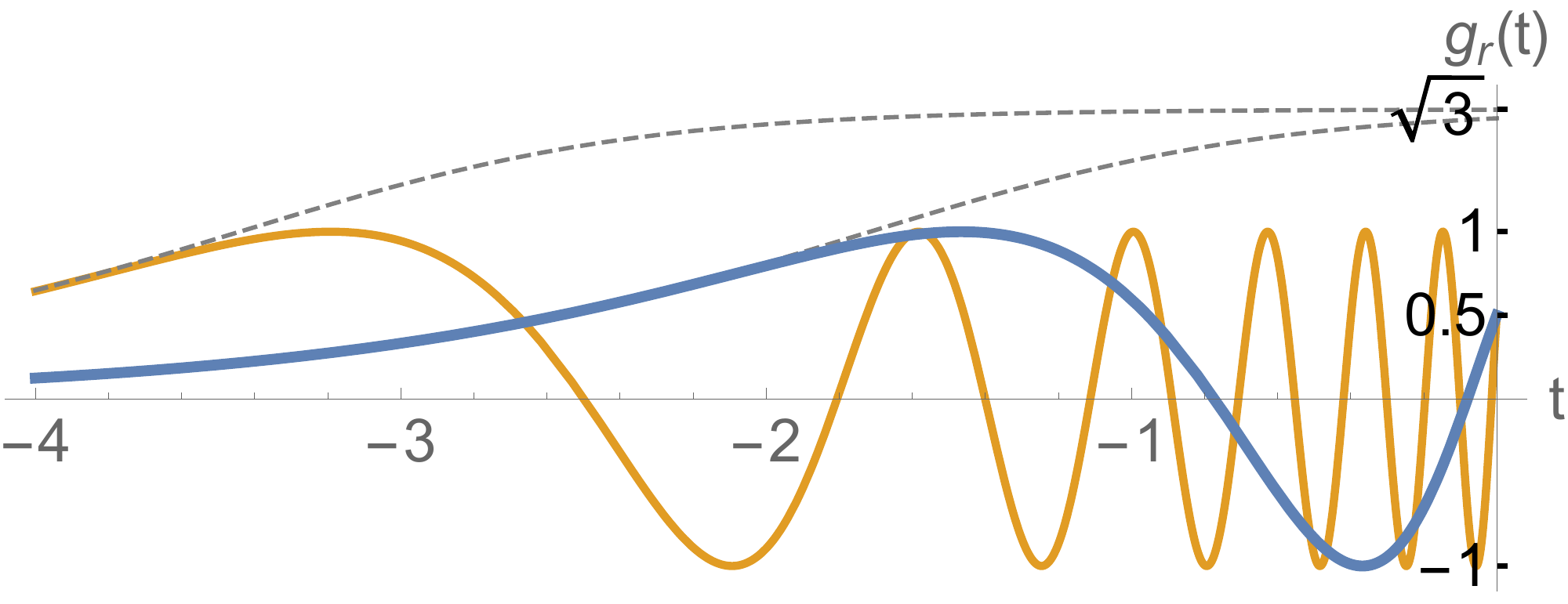}
	\caption{\label{FIG:BANOR}   {Coupling trajectories as a function of RG time. Both exact solutions (solid lines) yield the same renormalised coupling $g_r({\sf 0})=1/2$, but come from different branches of (\ref{gren}) corresponding to $g_0 = \pi/6+2\pi$ and $g_0=\pi/6+12\pi$. The 1-loop trajectories (dashed lines) tend to the fictitious fixed point at $g_r=\sqrt{3}$.}}
\end{figure}

The coupling trajectories are compared in Fig.~\ref{FIG:BANOR}. Those calculated from the $1$-loop $\beta$-function deviate from the exact solutions when $g_r$ first approaches one, and then tend to the fictitious fixed point at $g_r=\sqrt{3}$. Depending on initial conditions, a given exact solution can exhibit arbitrarily many  {rapid oscillations, corresponding to passing through many branches of the $\beta$-function. This behaviour is representative of the chaos underlying the flow. To make this explicit, observe that if we had used the probability $|A_0|^2$ in our renormalisation condition, rather than the amplitude, we would have studied the flow of $x(\rt):=g_r^2(\rt) = \sin^2 g_0 e^\rt$. This is (up to a trivial rescaling of $\rt$) well-known as the function which interpolates the chaotic behaviour of the discrete logistic map with parameter $4$~\cite{Schro,Curtright2010,Curtright2011}.}  {Now, following~\cite{Curtright-RG}, define}  {the ``$c$-function'' by $\ud c(g_r)/\ud g_r = \beta(g_r)  \implies \ud c(\rt)/\ud\rt = \beta^2(g_r(\rt))$.}
As a function of the coupling, $c$ is multivalued, but as a function of $\rt$ it is clearly monotonic.  Thus  {we find that} the JCM provides a simple, physical complement to the examples in~\cite{Curtright-RG} showing that, contrary to what may be inferred from the $c$ or $a$-theorems~\cite{Zamolodchikov:1986gt,Cardy:1988cwa,Anselmi:1997am,Komargodski:2011vj,Nakayama:2013is}, chaotic coupling trajectories are not ruled out by the existence of monotonic flow functions.

 {We note that the RG flow of the coupling in the Ising model with imaginary magnetic field~\cite{Curtright2011,Curtright-RG}, is also described by the logistic map with parameter~4~\cite{Dolan1995-1,Dolan1995-2,Curtright2010}. Interpolating the dynamics and renormalisation of such discrete systems through continuous functions following~\cite{Schro} allows an interpretation of RG flows in terms of (quasi) Hamiltonian dynamics~\cite{Curtright2010,Curtright2011}. For the logistic map, the interpolating function is our $x(\rt)=g_r^2(\rt)$. As such it is intriguing to note that the RG flow of the JCM is shared with that of the Ising model: for the flow of $x(\rt)$ corresponding to that in Fig.~\ref{FIG:BETA} see~\cite[Fig.~2]{Curtright2010}. These same structures arise, though, through quite different mechanisms, as we now discuss.}

\paragraph*{(Lack of) periodicity.}
For an RG trajectory which flows around a closed loop, a limit cycle, the parameters in the Hamiltonian return to their original values after a finite RG time~\cite{LeClair2004},  {so the theory} is periodic as a function of the flow~\cite{LeclairRussianDoll}. Special Circumstances are required for such behaviour, see e.g.~\cite{LECLAIR2004407}, but the situation here is somewhat different. The $\beta$-function (\ref{min-beta}) describes the variation of the coupling with respect to time, \textit{not} with respect to integrating out modes in a Wilsonian approach. It is defined in terms of the $S$-matrix, and as such inherits its periodic features from those of time evolution in the JCM. Periodicity is thus present \textit{in this sense}. However (and even without considering the $\beta$-function), different branches of solution yield \textit{different} physics, as shown in Fig.~\ref{FIG:SPEK}. As a result, a single measurement is not enough, non-perturbatively,  to determine the single coupling in the JCM. This is markedly different to what happens in perturbation  {theory, but we can explain it} as follows. The JCM interaction $V$ can only change the photon content of an initial number state $\ket{n}$ by $\pm 1$. Hence, the theory splits into a product of decoupled \textit{two-level} photon subsytems~\cite{Stenholm,Shore1993,Bina2012,Greentree2013}.  
	In our example, $n=0$ and all amplitudes in the sub-system (transitions between superpositions of $\ket{0,\uparrow}$ and $\ket{1,\downarrow}$) are indeed periodic as $g_0$ changes from one branch to another. However, the same periodicity does not extend to the entire spectrum of probabilities $\mathbb{P}_j$ because the renormalisation condition is essentially blind to all other, decoupled, subsystems.

\paragraph*{Effective $S$-matrix.}
%
In the `average effective action' approach to the RG, see~\cite{Gies:2006wv,Delamotte:2007pf}, one constructs a function depending on a flow parameter $k$ which, for $k \to\infty$, reproduces the classical action of the theory, and for which a change $k\to k - \delta k$ corresponds to integrating out quantum fluctuations with energy in the range $(k-\delta k,k)$. As such the full effective action of the quantum theory is recovered as $k\to 0$.   {Here we construct}  {a flow from the `bare' interaction Hamiltonian $V$ of the JCM, to the full $S$-matrix of the theory}.  We do so by re-introducing the detuning $\omega_a-\omega$ (the gap between the photon and atomic energies) and using it as a flow parameter. As before, we switch off the coupling at time $T$, so $U(T)$ is the $S$-matrix. Define $k= T(\omega_a-\omega)/2$, for $0 \leq k <\infty$, and write $U_k$ for the $S$-matrix with this detuning. For large $k$, $U_k$ behaves as  {(see Appendix~A and B)} 
\be\label{UK1}
	U_k  \sim \cos k - 2i \tau_3 \sin k -i g_0 V \frac{\sin k}{k} \;,
\ee
which shows that \textit{transitions} between atomic levels are suppressed as $1/k$ because of the large energy gap $\propto k$ between them. Thus the detuning acts as a mass scale which suppresses quantum fluctuations. Subtracting the diagonal contributions in (\ref{UK1}), $U_k$ is clearly proportional to the bare vertex $g_0 V$ up to a factor. Given this, we define an \textit{effective transition matrix} $\mathcal{T}_k$ by
\be\label{flow-def}
\begin{split}
	\mathcal{T}_k = U_k  - \big(\cos k - 2i \tau_3 \sin k\big) -i g_0 V(1 - \frac{\sin k}{k}\big) \;,
\end{split}
\ee
which obeys the two limits
\be
\label{graenser}\begin{split}
	\mathcal{T}_k  &\to  -i g_0 V \;, \quad \quad\quad k \to \infty\;, \\
	\mathcal{T}_k  &\to \mathrm{e}^{-i g_0 V} - 1 \;, \,\quad k \to 0\;.
\end{split}
\ee
Hence $\mathcal{T}_k$ interpolates between the bare vertex in the~UV, $k\to\infty$, where all quantum transitions are suppressed, and the \textit{T}--matrix proper (the $S$--matrix minus the forward scattering contribution) in the IR, as $k\to 0$ and the suppression is removed. This is our Hamiltonian analogue of the average effective action.  {Nicely, the subtractions in (\ref{flow-def}) mirror the usual subtraction of regulator-function dependent terms from the average effective action~\cite{Berges:2000ew}, which ensures the correct UV behaviour in~(\ref{graenser}).}
%
\begin{figure}[t!!!]
	\includegraphics[width=\columnwidth]{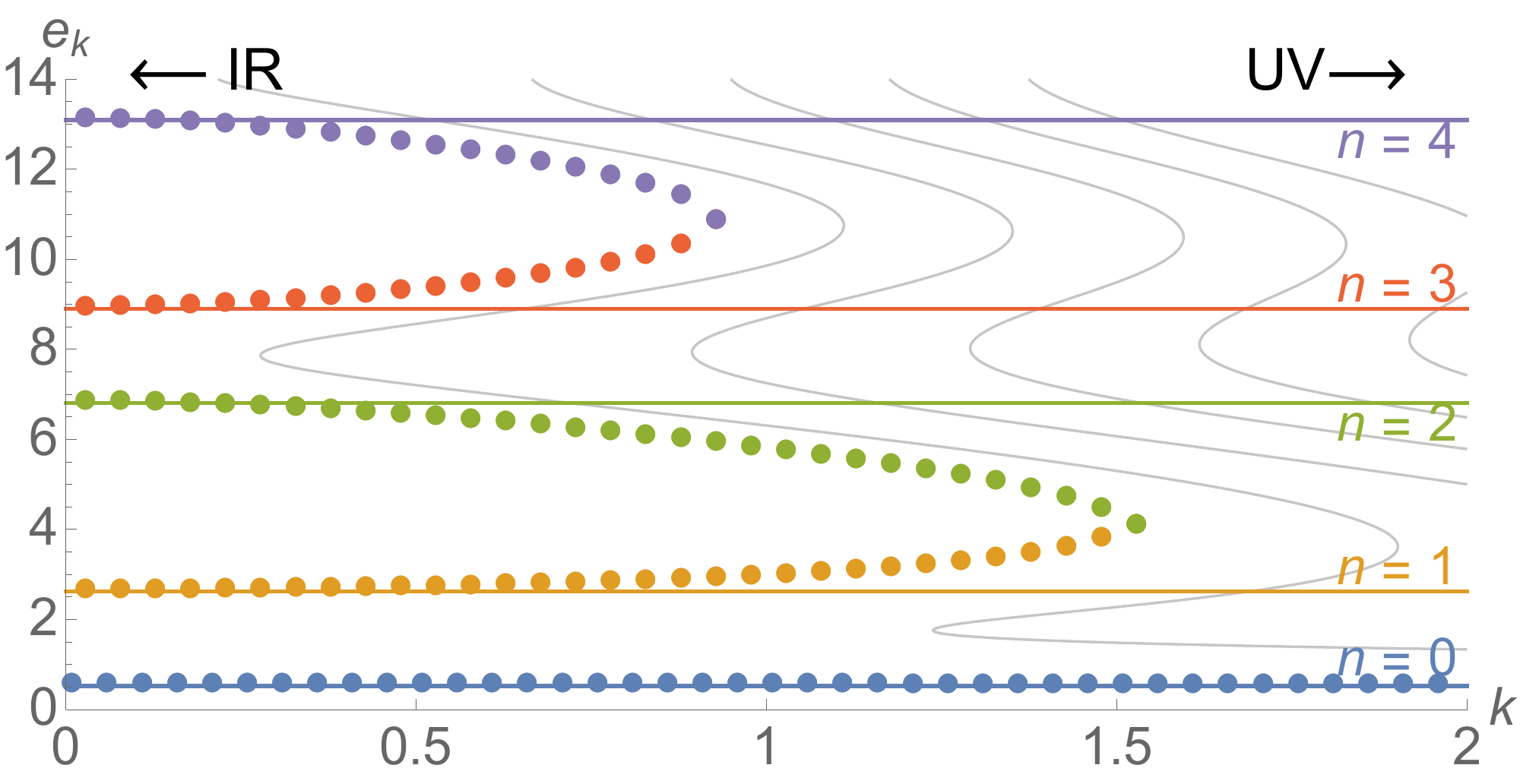}
		\caption{\label{FIG:BIR} Solutions to the renormalisation condition (\ref{ren-def-EL}) defining the coupling $e_k$, at $g_r=1/2$. Points show numerical solutions. As $k$ decreases toward 1, multiple solutions appear in the flow, and interpolate toward the IR solutions (\ref{gr-sol}) (horizontal lines) shown for $n=0\ldots 4$.
		 }
\end{figure}

 {Note that the effective $T$-matrix} (\ref{flow-def}) is a single-valued function of $k$, and does not exhibit any periodicity. Can the structures seen above, in particular the multi-valued bare coupling, then reappear?   {To answer this, we let the coupling become $k$-dependent, writing $g_0 \to e_k$, and adjust $e_k$ to preserve our renormalisation condition under the flow; here that condition is just $g_r = i \bra{1,\downarrow} \mathcal{T}_k \ket{0,\uparrow}$ which, for given, fixed $g_r$ is (see Appendix B):}
\be\label{ren-def-EL}
\begin{split}
	g_r= (1-\frac{\sin k}{k})e_k + \frac{e_k}{\sqrt{k^2+e_k^2}}\sin \sqrt{k^2 + e_k^2} \;.
\end{split}
\ee
In the IR, $k \simeq 0$, we can solve this immediately to find
\be\label{gr-sol}
	g_r \simeq \sin e_k \implies e_k \simeq \arcsin_n g_r \;,
\ee
which is just the multi-valued solution of the renormalisation condition above. In the UV, we must solve (\ref{ren-def-EL}) numerically; the solution(s) are shown in Fig.~\ref{FIG:BIR} along with the exact IR result (\ref{gr-sol}). For large $k$, there is only a single solution to (\ref{ren-def-EL}), which deviates only slightly, for $k> 0$, from the all-orders perturbative solution $e_k = \arcsin g_r$ (the $n=0$ branch). As $k$ decreases toward the IR, however, additional solutions appear  {in pairs in the flow. As $k\to 0$ these interpolate to the $n^\text{th}$ and $(n+1)^\text{th}$ branch solutions for $n\geq 1$, as shown,} with the higher $n$ solutions appearing at smaller $k$. Thus, the perturbative solution to the renormalisation  {condition exists} for all~$k$, while the `non'-perturbative solutions appear at \textit{finite RG times}.  {We find therefore that, in the IR}, there are several possible endpoints of the flow, recovering the multiple different couplings $\arcsin_n g_r$. The faint grey lines in Fig.~\ref{FIG:BIR} show contours of (\ref{ren-def-EL}) for constant $g_r$; these illustrate that for $g_r>1$ solutions to the renormalisation condition can appear and disappear in the flow as $k$ decreases, but that they never reach the IR at $k=0$; this is consistent with the fact that there no \textit{physical} solutions to the renormalisation condition in the IR for $g_r>1$.

\paragraph*{Discussion.}
 {We have shown that renormalisation of the apparently simple Jaynes-Cummings model provides an exactly solvable example of a theory with exotic coupling trajectory and multi-valued beta function.}   {One consequence of these structures is that, non-perturbatively, there can be multiple solutions of a given renormalisation condition, meaning that a single measurement is not enough to determine the single free parameter (the coupling) in the theory.  {(See Appendix C for a discussion of multiple measurements.)}}  {This has physical consequences. For example, a misidentification of the coupling, or limiting the coupling to the perturbative branch, could mean missing the famous collapse-revival physics of the JCM~\cite{Cummings:1965zz,Eberly:1980zz}, which is a strong-coupling phenomenon~\cite{Nazir1,Ekman:2020vsc}.}

  {The question arises of how our results would be affected by the addition of more structure, which could extend both the theoretical and experimental applicability of the model. The RG structure of field theories like QED is extremely rich~\cite{Gockeler:1997dn,Gies:2004hy,Gies:2020xuh}, so a natural first step beyond our results would simply be to add the `counter-rotating' terms to the Hamiltonian which turn the JCM into the Rabi model. The fact that the Rabi model is also solvable~\cite{Braak} offers scope for progress here.}

 {We have also introduced an $S$-matrix analogue of the average effective action, using the detuning of the JCM as a flow parameter. Normally one can find an exact flow \textit{equation} for the average effective action, but not solve it exactly. In the JCM we can essentially jump straight to the exact solution. It would nevertheless be interesting to investigate the corresponding flow \textit{equations} (which can be set up for (\ref{flow-def}) or (\ref{ren-def-EL}) by taking the derivative with respect to the detuning) in order to explore how the bifurcations in Fig.~\ref{FIG:BIR} arise; this could give insight into RG flow equations in other theories. Certainly the essential idea of an effective $S$-matrix is not limited to the JCM, and so may offer an alternative approach to Hamiltonian RG studies~\cite{Vacca:2012vt} in other theories.}

 {Our investigation highlights the dangers of inferring results from perturbative renormalisation (even in simple settings stripped of the complications of removing divergences) and provides non-perturbative insight into the RG in a physical and accessible setting.}

\begin{acknowledgments}
\textit{AI thanks Holger Gies for stimulating discussions and useful comments on a draft of this manuscript. AI is supported by the Leverhulme Trust, grant RPG-2019-148.}
\end{acknowledgments}

\appendix

\section*{Appendix A: arbitrary detuning}
Let $\Delta=\omega_a-\omega$ be the difference between the atomic and photon energies, or `detuning'. The time-evolution operator $U(t)$ for arbitrary detuning is~\cite{Stenholm}
\be\begin{split}\label{UU}
	U(t) = &\Big[\cos t \psi - i \frac{\Delta}{2}\frac{\sin t \psi}{\psi} \Big]\kett{\uparrow}\brat{\uparrow} -i g \frac{\sin t \psi}{\psi} a\tau_\LCp \\
	+&\Big[\cos t \phi + i \frac{\Delta}{2}\frac{\sin t \phi}{\phi} \Big]\kett{\downarrow}\brat{\downarrow}   -i g \frac{\sin t \phi}{\phi} a^\dagger \tau_\LCm \;,
\end{split}\ee
in which $\phi = \sqrt{g^2 a^\dagger a + \tfrac{\Delta^2}{4}}$ and $\psi = \sqrt{g^2 a a^\dagger + \tfrac{\Delta^2}{4}}$.

\section*{Appendix B: large detuning}
The dimensionless variables used to construct the Wilsonian flow of the effective $S$-matrix are $e_k = g t$ and $k=t \Delta/2$ for some fixed $t$. Defining $\varphi = \sqrt{e_k^2 a^\dagger a + k^2}$ and $\vartheta = \sqrt{e_k^2 a a^\dagger + k^2}$, the time evolution operator (\ref{UU}) in terms of the new variables is
\be\begin{split}
	U_k := &\Big[\cos \vartheta - i k \frac{\sin \vartheta}{\vartheta} \Big]\kett{\uparrow}\brat{\uparrow} -i e_k \frac{\sin \vartheta}{\vartheta} a\tau_\LCp \\
	+&\Big[\cos \varphi + i k \frac{\sin \varphi}{\varphi} \Big]\kett{\downarrow}\brat{\downarrow}  -i e_k \frac{\sin \varphi}{\varphi} a^\dagger \tau_\LCm \;.
\end{split}\ee
From this we can read off the large-$k$ behaviour. Using
\be\begin{split}
	\Eins &= \kett{\uparrow}\brat{\uparrow} +\kett{\downarrow}\brat{\downarrow} \;, \quad 
	2\tau_3 = \kett{\uparrow}\brat{\uparrow} -\kett{\downarrow}\brat{\downarrow} \;,
\end{split}
\ee
we find directly that, for $k\to \infty$, 
\be\label{limUK}
	U_k  \sim \cos k - 2i \tau_3 \sin k -i e_k \big(a^\dagger \tau_\LCm + a \tau_\LCp\big) \frac{\sin k}{k} \;,
\ee
as used in the text. For this limit to hold we must, strictly, place a cutoff $\Lambda$ on the allowed mode occupation number (which is otherwise unbounded). Hence $V$ should be considered as projected onto a finite subset of modes throughout the effective-S-matrix calculation. This does not affect the results: $\Lambda$ is simply analogous to a UV cutoff, to be removed at the end of the calculation.

\section*{Appendix C: multiple measurements}
%
\begin{figure}[t!!!]
	\includegraphics[width=0.48\columnwidth]{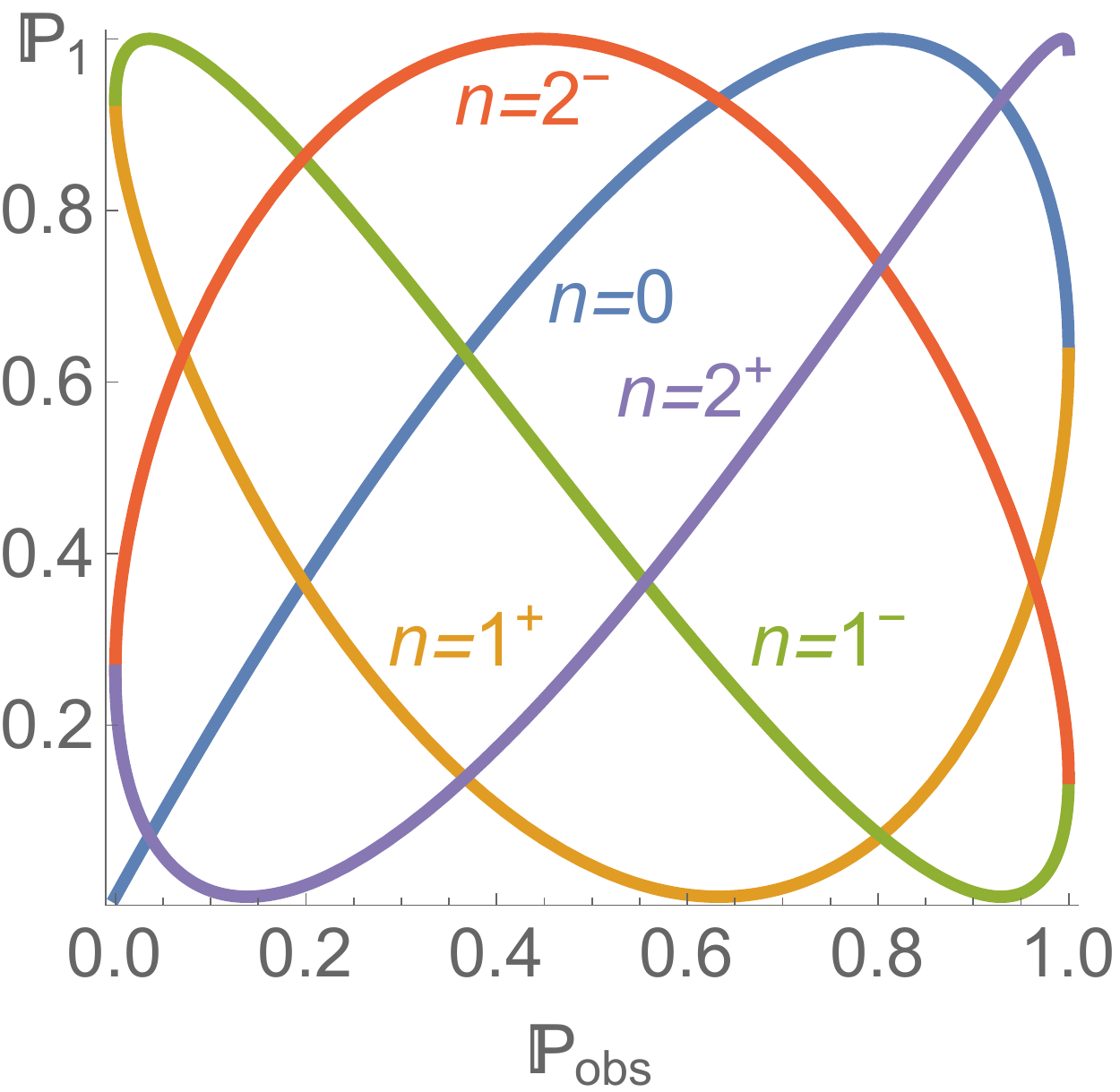}
	\includegraphics[width=0.45\columnwidth]{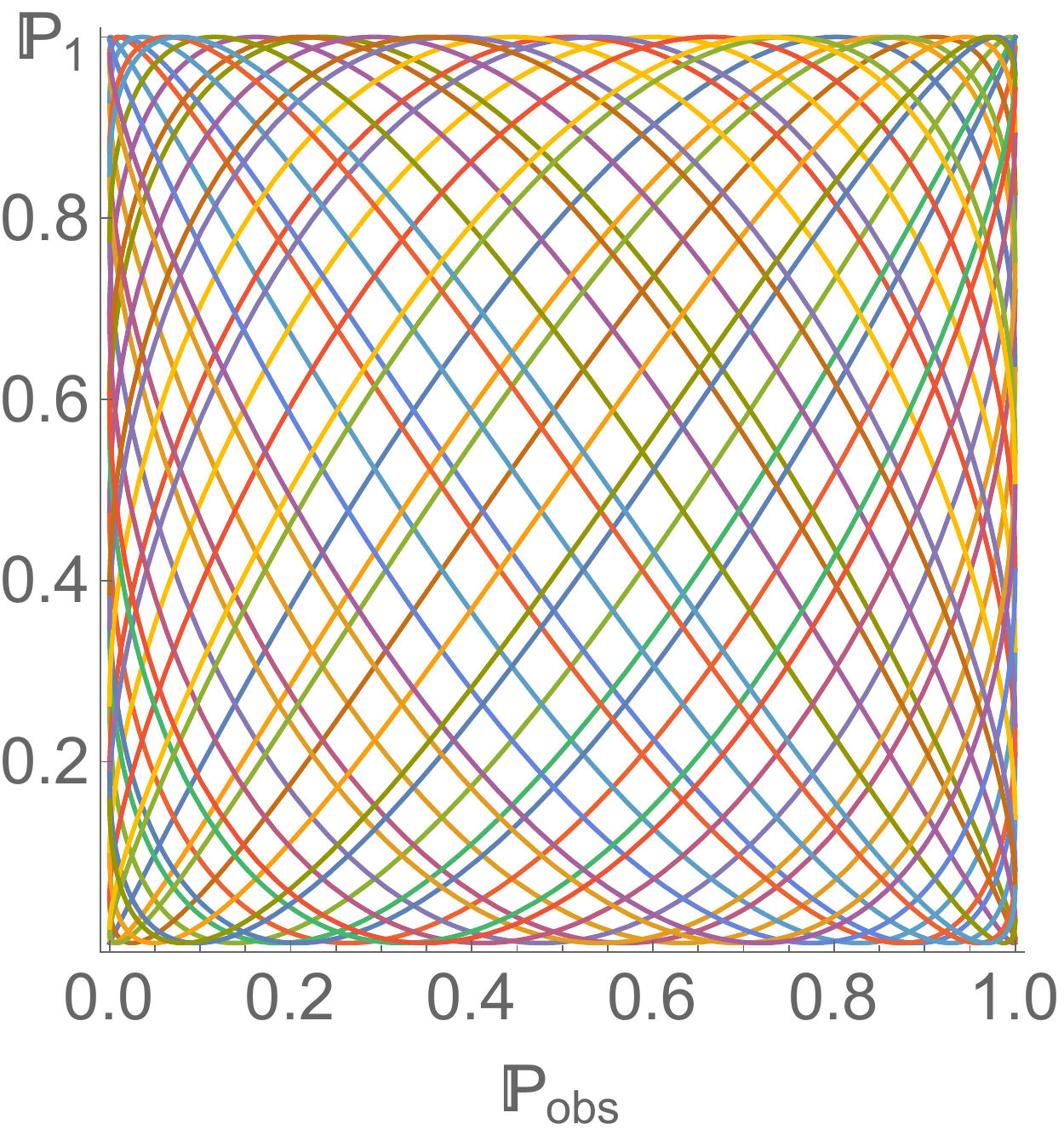}
	\caption{\label{FIG:PP1} The relationship between $\mathbb{P}_\text{obs}$ and $\mathbb{P}_1$ for different branches (\ref{gren2}) with $\pm$ labelling the sign of square root. The first 20 of each sign are shown on the right.}
\end{figure}
In the text, we used a measurement of a particular scattering probability at a given time as a renormalisation condition to fix the coupling of the JCM. This is analogous to using a scattering probability (or amplitude) at fixed energy scale to renormalise the coupling in QFT. Using any single transition amplitude would have led to the same conclusion (as the form of (5) in the text shows), namely that a single measurement is not enough, non-perturbatively, to determine the coupling. We discuss here the extent to which a second measurement can be used to uniquely fix the coupling of the JCM model.

Suppose then that we have measured $\mathbb{P}_\text{obs}$ for $\mathbb{P}_0$, as in the text, and identified the bare coupling $g_0$ in terms of the $n^\text{th}$ branch of arcsin as
\be\label{gren2}
	g_0 =  n \pi + (-1)^n \arcsin \big(\pm \sqrt{\mathbb{P}_\text{obs}}\big) \;.
\ee
Suppose also that we have a second measured value for some other process, say, for $j>0$,
\be
	\mathbb{P}_j = |  \bra{j+1,\downarrow} U(t) \ket{j,\uparrow}  |^2 = \sin^2 g_0 \sqrt{j+1} \;.
\ee
Note that if $\sqrt{j+1} \in \mathbb{Z}$ then $\mathbb{P}_j$ is independent of the choice of branch in (\ref{gren2}), and we gain nothing. Hence we must pick $j$ such that $j+1$ is not a perfect square; note that $\sqrt{j+1}$ is then irrational (Theaetetus' theorem), which is key to what follows. Taking $j=1$ to illustrate, Fig.~\ref{FIG:PP1} shows how different branches (\ref{gren2}) yield different predictions for $\mathbb{P}_1$ at a given $\mathbb{P}_\text{obs}$. As the situation is clearly complex when all possible branches are considered, it is helpful to take a specific example. 

Consider a scenario in which the excited state of the atom $\ket{0,\uparrow}$ is unstable and always decays. Then the observed decay probability is unity, $\sin^2 g_0 = 1$, and so we may write
\be
	g_0 = \big(n + \tfrac{1}{2}\big)\pi\;, \quad n \in \mathbb{Z}.
\ee
We can restrict to $g_0 \geq 0$, hence $n\geq 0$, since observables in the JCM are invariant phase transformations $a\to e^{i\phi}a$ of the photon ladder operators, which for $\phi=\pi$ is equivalent to $g_0\to -g_0$.
The question to answer is whether or not a given value of some $\mathbb{P}_j$ is enough to determine the branch $n$, and fix the coupling, uniquely. The equality of $\mathbb{P}_j$ for \textit{two} branches $n$ and $n'$ would require
\be
\begin{split}
	\sin^2 \sqrt{j+1}(n+\tfrac12)\pi = \sin^2 \sqrt{j+1}(n'+\tfrac12)\pi \;.
\end{split}
\ee
The solution to $\sin^2a = \sin^2b$ is just that one of $a\pm b$ differ by an integer multiple of $\pi$; hence we have either
\be
	\sqrt{j+1}(n-n') \in \mathbb{Z} \;,
\ee
for which the only solution is $n=n'$ precisely because we have taken $\sqrt{j+1}$ irrational, or
\be
	\sqrt{j+1}(n+n'+1) \in \mathbb{Z} \,,
\ee
which similarly has no solution for $n,n'\geq 0$. Here, then, a second measurement of $\mathbb{P}_1$ is in this case enough to determine the coupling.

More generally, there is a subtle dependence on the irrationality (or otherwise) of the arguments in the various trig functions appearing. This is a direct consequence of the chaos in the underlying logistic map, which our functions interpolate~\cite{Dolan1995-1}: as is well known, whether the equivalent parameters are rational or irrational determines the (non) periodicity of the sequence generated by the map~\cite{Schro}.


\begin{thebibliography}{44}
\bibitem [{FN2()}]{FN2}%
  \BibitemOpen
  \href@noop {} {}\bibinfo {note} {Functions of operators are defined by their
  power series.}\BibitemShut {Stop}%
\end{thebibliography}
\end{document}